# Phase Transition Induced Fission in Lipid Vesicles

C. Leirer[1,2], B. Wunderlich[1], V. M. Myles[1], M.F. Schneider[1*]

[1] University of Augsburg, Experimental Physics I, D-86159 Augsburg, Germany
[2] Nils Bohr Institute, Copenhagen, DK

*Corresponding author:*
*Matthias F. Schneider*
*Phone: +49-821-5983311, Fax: +49-821-5983227*
*matthias.schneider@physik.uni-augsburg.de*
*University of Augsburg, Experimental Physics I,*
*Biological Physics Group*
*Universitätstr. 1*
*D-86159 Augsburg, Germany*

**Abstract**

In this work we demonstrate how the first order phase transition in giant unilamellar vesicles (GUVs) can function as a trigger for membrane fission. When driven through their gel-fluid phase transition GUVs exhibit budding or pearl formation. These buds remain connected to the mother vesicle presumably by a small neck. Cooling these vesicles from the fluid phase ($T>T_m$) through the phase transition into the gel state ($T<T_m$), leads to complete rupture and fission of the neck, while the mother vesicle remains intact. Pearling tubes which formed upon heating break-up and decay into multiple individual vesicles which then diffuse freely. Finally we demonstrate that mimicking the intracellular bulk viscosity by increasing the bulk viscosity to 40cP does not affect the overall fission process, but leads to a significant decrease in size of the released vesicles.

**Introduction**

Budding, fission, and fusion of biological membranes are essential mechanisms for intracellular trafficking and the transport of nutrients and waste across the cell membrane. The underlying physical principles, however, still lack a fundamental understanding. Membrane mechanics have been shown to be of major importance and can be controlled by both lipid and protein characteristics and behaviour [1, 2]. The changes in shape seen during membrane trafficking and the morphology of lipid vesicles have been shown to be determined mainly by the minimization of bending energy for a certain bending modulus and area to volume ratio [3]. Theoretical studies by Chen et al. have shown that membrane fission occurs when the minor lipid preferentially remains at regions of high positive Gaussian curvature [4], an effect which is enhanced near a critical state. Further, Roux et al. demonstrate that phase separation in long multicomponent membrane tubes supports the process of fission [5, 6]. In contrast to our studies, these authors did not investigate the gel-fluid phase transitions in lipid vesicles.

Here we describe how heating through the lipid phase transition, in pure phospholipid membranes, can be used to produce membrane structures connected with a narrow neck which subsequently break off upon cooling into or below the phase transition regime. While the fission process is not qualitatively affected by the viscosity of the medium, the size scale of the resulting individual vesicles decreases an order of magnitude from ~ 5µm to ≤ 0.5µm when the bulk viscosity is raised to 40cP [7]. This value is well within the range of intracellular microviscosities which are reported to be between 1cP and 140cP [8-11]. Recalling that many biological membranes are close to a phase transition [12, 13], we propose that the lipid phase transition may be a relevant factor for vesicle fission in biological systems.

**Materials and Methods**

Dipalmitoylphosphocholine (DPPC) and Dimyristolphosphocholine (DMPC) dissolved in chloroform (20mg/ml) were purchased from Avanti Polar Lipids (Alabaster, Alabama, USA) and were used without further purification. Vesicle electroformation was performed as described elsewhere [14-16]. All aqueous solutions were prepared with ultrapure water (pure

Aqua, Germany) with a specific resistance of 18.2MΩ. For vesicle preparation, we spread a small amount of lipid in chloroform on an Indium Tin Oxide (ITO) coated glass substrate which was then placed in a vacuum to remove any traces of organic solvent. The swelling chamber was formed using a Teflon spacer to separate two such ITO slides. Subsequently, an AC field of 1mV/mm and 10Hz was applied between the conducting ITO slides and the assembled chamber was fixed and placed in a 50°C thermal water bath. The electroswelling process was completed after 3 hours and Giant Unilamellar Vesicles (GUVs) up to 100μm in diameter were harvested for use in experiments. Fluorescence microscopy of the vesicles was performed with the addition of 0.1% fluorescently labelled DHPE (Texas Red from Invitrogen, USA). Experiments were performed in a temperature controllable chamber with optical access from the bottom and top and the temperature was controlled with the aid of a standard heat bath (Julabo).

**Results**

*Phase transition induced morphological changes*

When heating the vesicle shown in Fig. 1a, from a starting gel-like state ($T = 28°C$) to a point above ($T = 45°C$) its transition temperature $T_m = 35°C$ excess area (~20%) [17] is created which must be accommodated for by a new morphology. This, in turn, is determined by minimizing the free energy of bending for the experimental boundary conditions. Such transitions have been studied extensively and entire phase diagrams have been published for mechanically stable configurations as a function of the reduced volume $v_r$ as well as the area difference between the two monolayers ($\overline{\Delta a_0}$) [18, 19]. Here, $v_r$ represents the deviation of the area/volume ratio ($A/V$) from the spherical shape ($v_r = 1$). Since the temperature quench into the fluid phase takes place in ~1s, it does not provide enough time for a significant volume exchange. Therefore, the new reduced volume $v_r$ can be calculated from the increase in area per molecule upon heating (at constant volume) to $v_r \approx 0.75 \pm 0.1$. Consulting the phase diagram of Doebereiner [20], the budded topology as observed in Fig. 1a indicates that $\overline{\Delta a_0} \approx 2.0$ (Fig. 1b). It is important to note that it was not possible to predict the final shape of the vesicle, i.e. whether extravesicular or intravesicular (Fig. 2a) budding was going to occur. We believe this is a consequence of variations in $\overline{\Delta a_0}$ which happen during preparation and cannot be easily controlled experimentally. In roughly 80% of all of our experiments either extravesicular or intravesicular budding occurred (Fig. 1a and 2a) while the remaining 20% exhibited no significant morphological transition. Analyzing the internalized buds with respect to the area we find an area increase of $21 \pm 5\%$ as expected from melting. Importantly, the two vesicles did not disconnect during the entire budding process, but remained attached to their mother vesicle presumably by a small neck or tether.

*Fission of buds*

After equilibration in the fluid phase ($T = 45°C$) we cooled the vesicle with the internalized buds rapidly (1°C/s) to a point below its phase transition temperature ($T = 28°C$), which lead to the break-up and expulsion of the multiple internalized vesicles (Fig. 2b). Since the expulsion process has been described in detail elsewhere [21], we here focus on the actual separation or fission process of membrane segments. In Figure 3a, fluorescent images of "pearling" vesicles in the fluid phase are shown. Upon cooling through the lipid phase transition, these "pearls" begin to straighten out or reintegrate (image 2 of Fig 3a) and eventually separate into individual vesicles (image 3 and 4 of Fig 3a) which diffuse freely. The addition of an incorporated fluorphore (Texas-Red) indicates that phase transition and fission are correlated. The small black domains which appear where the pearling tubes begin to disconnect (see arrows in Fig. 3a image 2) indicate the coexistence of the two phases. The correlation of the onset of break-up and gel-like domains (both appear simultaneously within

±1s precision) point to the phase coexistence as a possible driving force for membrane fission. This is in contrast to the work of Roux et al. [5], which observed fission in the presence of cholesterol without gel/fluid phase coexistence.

As mentioned, the experiments displayed in Fig. 3a have been performed at very high heating rates of 1°C/s. In order to study the importance of dynamics for the fission process, the experiments were repeated not only at a lower cooling rate of 10°C/min, but also at an elevated viscosity of 40cP, which is well within the range of reported intracellular microviscosities [8-11]. As demonstrated in Fig. 3b, we see that viscosity and cooling rate do not prevent fission to take place yet they strongly influence size scales, an effect which will be addressed in a separate manuscript in further detail [22]. In brief, the total energy to be minimized during budding consists of an elastic term (the bending energy) and a dissipative contribution (membrane and bulk viscosity) [7]. While the elastic term favors one large bud, the viscous term is minimized by multiple, smaller buds. Therefore, media viscosity favors the formation of vesiculated structures of reduced size which then break off during cooling.

Although a thorough theoretical treatment is beyond the scope of this experimental work, we would like to provide a brief phenomenological explanation of the fission process. We have recently shown that the conductivity of lipid membranes is increased during the lipid phase transition in which an increase in both pore probability and relaxation timescales is observed. A theoretical explanation has been presented based on an increase in area fluctuations $\langle \delta A^2 \rangle$ during phase transition according to [23]:

$$\langle \delta A \, \delta A \rangle = -k_B \left( \frac{\partial^2 S}{\partial A^2} \right)^{-1} = k_B T \bar{A} \kappa_T \qquad (1)$$

where $S$ is the entropy potential according to Einstein [24], $\bar{A}$ is the average area per lipid, $k_b$ the Boltzmann constant, $T$ the temperature and $\kappa_T$ the isothermal compressibility which exhibits a maximum in the phase transition regime [17] [25]. Taking numbers for $\kappa_T$ from literature [25], these area fluctuations are in the range of 1-20%. Such strong fluctuations in area will necessarily lead to pores in the lipid membrane. Microscopically, it is discussed that boundaries arising during the phase transition are the origin of these pores, a mechanism in accordance with Monte Carlo Simulations [26]. Along the same lines, it can be argued that the "flat" thermodynamic potential near $T_m$ reduces the line tension, as has been shown by McConnell [27]. Reduced line tension will, in turn, favour pore formation and membrane rupture (fission) and is therefore in agreement with the thermodynamic interpretation (Eq. 1) of our results.

In summary, we report that connected membrane networks separate from each other when cooling below $T_m$ takes place. We propose that strong thermodynamic fluctuations, which are maximized in the transition regime, create small pores [23] along the connecting neck essentially "perforating" the membrane and increasing the probability of total membrane rupture. Finally, we demonstrate that an increase in viscosity to 40cP results in sizes of ≤ 500nm even at low cooling rates. Taking into consideration reported microviscosities of ~100cP [9] and findings that, at least some, biological membranes are close to a phase transition [12, 13], we propose that the membrane phase transition is a suitable trigger for membrane fission that might be utilized by biological systems.

**Acknowledgments**

We would like to thank M. Kozlov for his stimulating comments. Financial support was provided by the Deutsche Forschungsgemeinschaft DFG (SFB 486, SPP 1313, SCHN 1077), the BMBF and the German Excellence Initiative via the "Nanosystems Initiative Munich (NIM)" and is gratefully acknowledged. MFS and CTL were supported by the Bavarian

Research Foundation. Further, we would like to acknowledge the help of A. Wixforth (Augsburg) for assistance during the preparation of this manuscript.

**Figure Captions**

Figure 1: (a) Image series of the phase transition induced separation of a GUV. (*1*) The system is equilibrated at 25°C then rapidly heated to ~45°C ($T_m = 35$°C) with a heating rate of 3°C/s. (*2*) 8 sec after heating the vesicle surface appears rather wrinkled due to the increased surface area. Before reaching its final shape (*8*), the vesicle relaxes through a series of (probably) metastable states including spherocylindric (*3,4*) and pear-like shapes (*5,6*). b) Mapping the final (stable) shape into the phase diagram (reproduced from [20]) reveals a reduced volume of ~ 0,76 and $\Delta a_o$ ~ 2,0.

Figure 2: Fluorescence images of DMPC/DPPC (1/1) GUVs ($T_m = 36$°C). (a) When heated from the gel (*T = 28°C*) to the fluid phase (*T = 45°C*) intravesicular buds appear, which however, stay connected with the outer membrane by a small tether. (b) Upon cooling the internalized vesicles are disconnected from the mother vesicle and are expelled. The area of the created buds was 21 ± 5% of the original mother vesicle.

Figure 3: Fluorescence images of DMPC/DPPC (1/1) lipid vesicles. (a) After heating from the gel (*T = 28°C*) to the fluid state (*T = 45°C*) vesicles remain connected in a long tube of pearls (image *1*). Cooling through the phase transition leads to break-up of the individual vesicles (image *3 – 4*). The onset of the break-up occurs simultaneously (within ~1s) with the appearance of gel-like domains indicated by the black patches in images *2* and *3*, taken 2 and 5 seconds after the first domains were detected (scan rate 10°C/s). After an additional 5s the vesicles were completely separated and diffused freely. The experiments have been performed on over 20 vesicles with the same outcome in the range of experimental accuracy (time

resolution ~ 100ms). (b) Increasing medium viscosity and decreasing the scan rate (10°C/min) results in size scales around and below optical resolution (≤ 500nm). However, the fission process itself is not affected and takes place as before when cooling the vesicle from 45°C (left image) to 28 °C (right image).

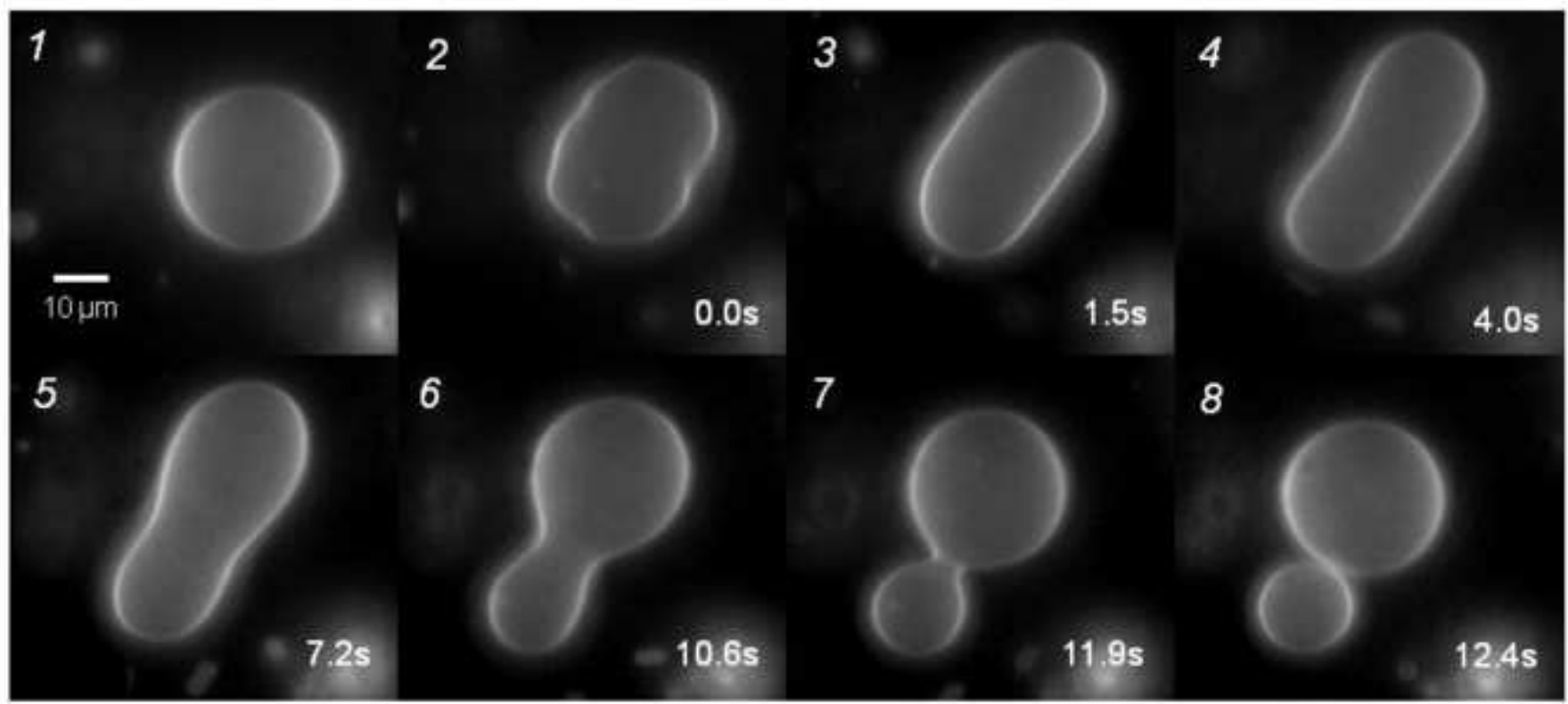

**Figure 1b**
**Click here to download high resolution image**

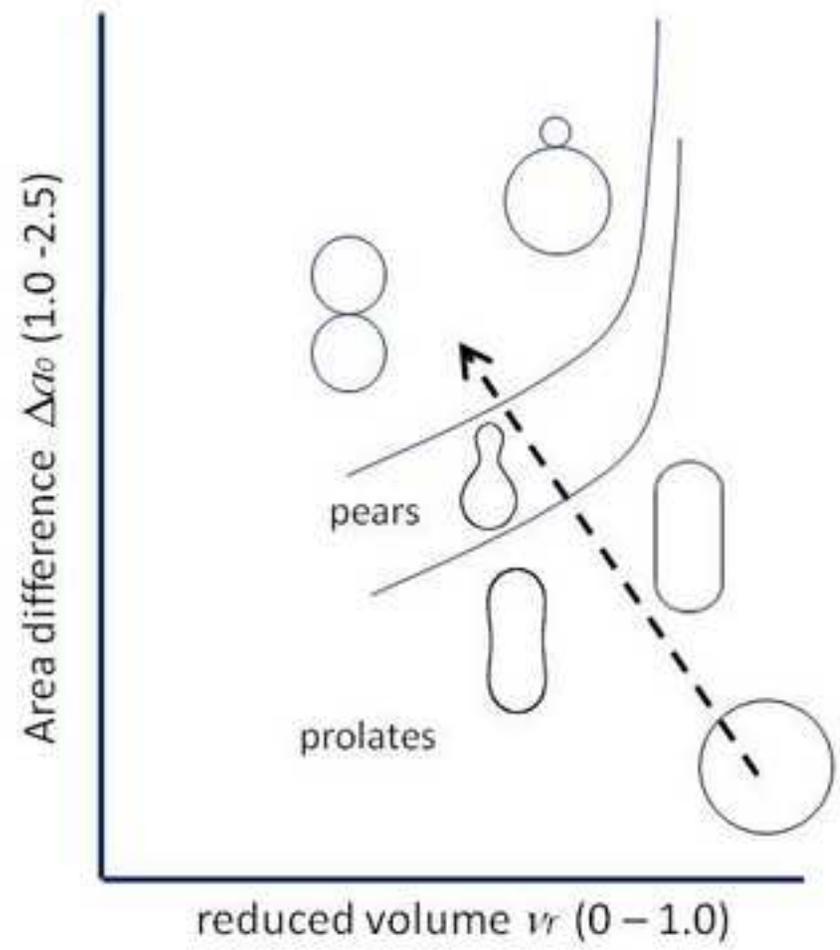

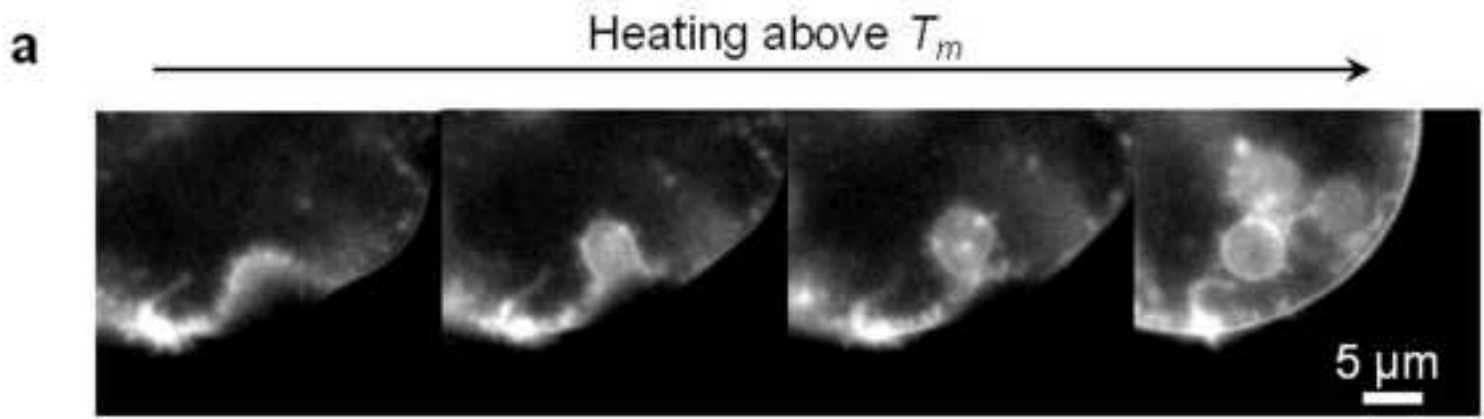

a — Heating above $T_m$

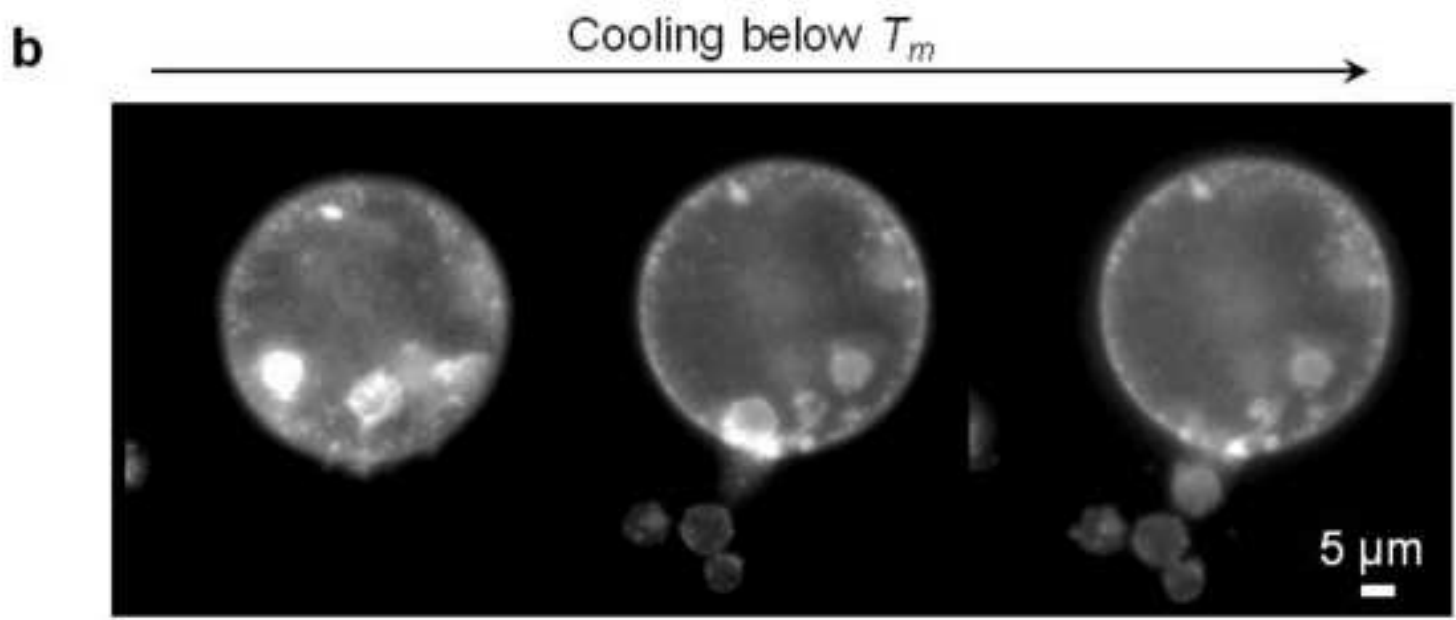

b — Cooling below $T_m$

**Figure 3a**
[Click here to download high resolution image]

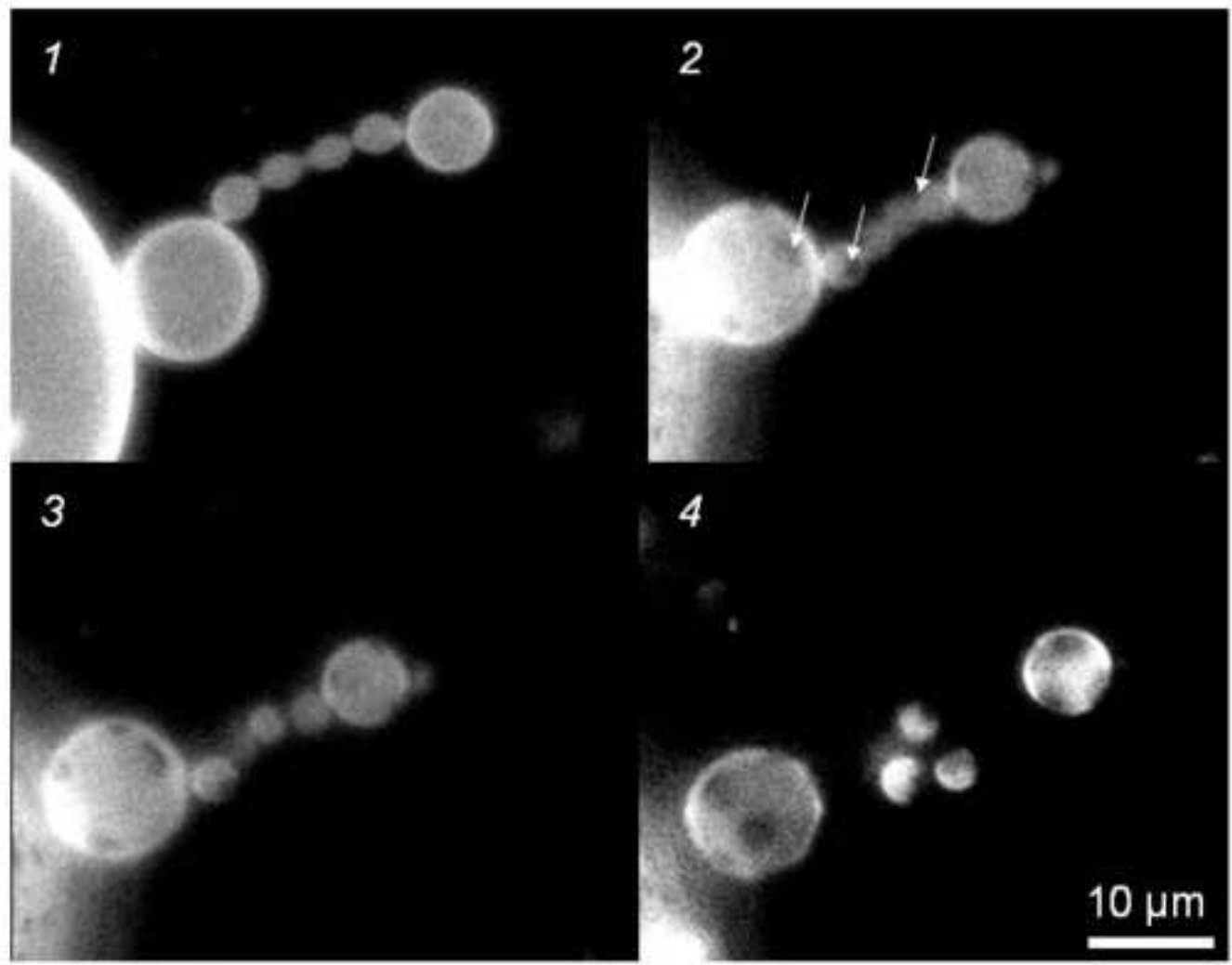

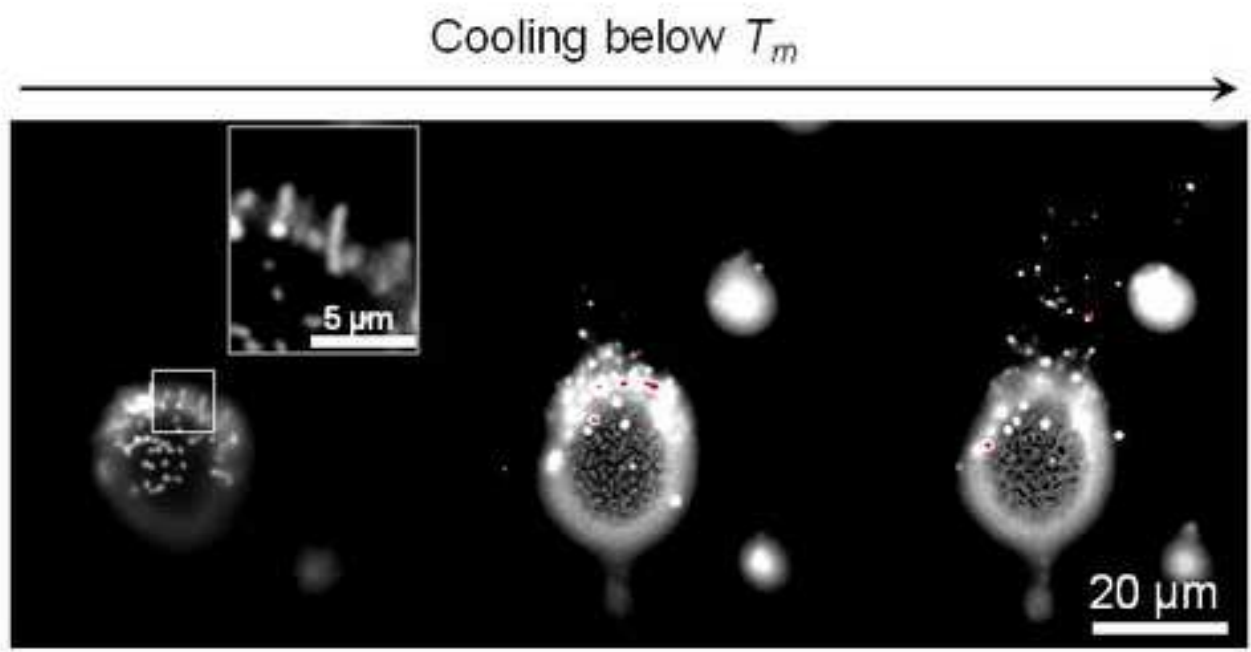